\documentclass[conference]{IEEEtran}
\usepackage{cite}
\usepackage{amsmath,amssymb,amsfonts}
\usepackage{graphicx}
\usepackage{url}
\usepackage{booktabs}

\begin{document}

\title{Implementation and Evaluation of NTT Arithmetic for ML-KEM on a CGLA}
\author{
\IEEEauthorblockN{Takuto Ando, Yasuhiko Nakashima}
\IEEEauthorblockA{Nara Institute of Science and Technology, Japan}
}

\maketitle

\begin{abstract}
FIPS~203 standardizes ML-KEM for post-quantum key establishment, and its polynomial multiplication relies on NTT butterflies with exact modular arithmetic over $q=3329$.
Dedicated NTT accelerators minimize latency with fixed modular arithmetic and stage schedules, whereas a CPU-Grounded Linear Array (CGLA) reuses one programmable linear datapath across several workloads.
Mapping the transform to this unchanged datapath requires exact FP32 reconstruction together with explicit stage transitions across the ARM--CGLA interface.
We implement an eight-stage cyclic radix-2 driver over the ML-KEM modulus by splitting each twiddle into 8-bit and 4-bit parts before modular reduction.
The evaluated driver differs from the standardized seven-layer incomplete negacyclic NTT and does not implement the full ML-KEM polynomial-multiplication path.
This sequence keeps every integer below $2^{24}$.
A 41-PE call fuses the first two radix-2 stages, and six 47-PE calls execute the remaining stages while scattering outputs into next-stage records.
Across four cohorts, 105 FPGA runs match all 817,152 output coefficients.
At batch~64, the measured FPGA end-to-end latency is 27.5\,$\mu$s per NTT, and the ASIC projection is 6.74\,$\mu$s.
With PE gating, the projected ASIC system energy is 10.1\,$\mu$J per NTT at batch~8 and 58.9\,$\mu$J at batch~64.
\end{abstract}

\begin{IEEEkeywords}
post-quantum cryptography, ML-KEM, number theoretic transform,
modular arithmetic, CGLA
\end{IEEEkeywords}

\section{Introduction}

With ML-KEM standardized as FIPS~203, post-quantum key encapsulation now requires implementations whose execution time and energy can be characterized on the target platform~\cite{fips203}.
Its polynomial multiplication repeatedly invokes the Number Theoretic Transform (NTT), whose butterflies require exact multiplication and reduction over $q=3329$.
An existing compute platform must therefore provide an exact NTT path alongside the non-cryptographic kernels that already occupy the same hardware.

Dedicated FPGA and ASIC accelerators address this requirement with modular multipliers, stage schedules, and memory structures built specifically for NTT execution~\cite{saoudi2024ntt,nguyen2024compact,gao2025bram}.
CGLA follows a different design point because it reuses one programmable datapath across several workloads.
Its 64-stage pipeline of processing elements (PEs) and local memory modules (LMMs) has already executed LLM, convolution, and matrix-vector kernels~\cite{uetani2024cgla,duong2025unet,kim2025knn}.
The objective of this work is to execute exact NTT arithmetic and carry its intermediate records across stages without adding an NTT-specific unit.
This objective is complicated by the coefficient-pair distance and twiddle assignment changing at every stage, unlike the regular dataflow of a matrix or convolution kernel.
Consequently, the mapping must preserve exact residues, fit the arithmetic dependency chain within 64 rows, and emit records that the following stage can consume directly.
Host work at each call boundary can otherwise dominate the cost of the mapped arithmetic.

We map NTT arithmetic over the ML-KEM modulus to the unchanged CGLA datapath and evaluate correctness, FPGA latency, and projected ASIC latency and system energy.
On the FPGA prototype, direct FP32 addition of partial products with different exponents produced one-ULP errors, so each product is reduced before recombination.
The first two stages are fused, and their output is repacked once to avoid repeated polynomial materialization and rearrangement across the ARM, DMA, LMM, and CGLA interfaces.
After this fused boundary, each CGLA call writes its output pair directly into the record layout consumed by the next stage.
The seven-call chain performs multiplication, modular reduction, butterfly correction, and record construction, while ARM NEON normalizes the records in place.
We then vary offload depth and batch size under a fixed FPGA boundary to measure configuration and transfer amortization and to project system energy.

Table~\ref{tab:design_space} summarizes representative implementations and states the basis of each reported metric.
\begin{table*}[t]
  \centering
  \caption{Representative NTT implementations and metric basis.}
  \label{tab:design_space}
  \footnotesize
  \setlength{\tabcolsep}{3.5pt}
  \begin{tabular}{p{0.12\textwidth}p{0.12\textwidth}p{0.07\textwidth}p{0.30\textwidth}p{0.29\textwidth}}
    \toprule
    Work & Platform & Prog. & Timing / energy basis & Implementation \\
    \midrule
    Saoudi~\cite{saoudi2024ntt} & Artix-7 FPGA & No & 0.4\,$\mu$s & dedicated NTT datapath \\
    Nguyen~\cite{nguyen2024compact} & FPGA & No & 1.10\,$\mu$s & pipelined NTT datapath \\
    Gao~\cite{gao2025bram} & FPGA & No & 40 cycles & streaming NTT/INTT datapath \\
    Zhu~\cite{zhu2026cim} & 28\,nm CIM & Partial & 10\,nJ & protocol-specific CIM array \\
    This work & CGLA & Yes & 27.5\,$\mu$s FPGA measured and 6.74\,$\mu$s ASIC E2E projected at batch~64, 10.1\,$\mu$J projected system energy at batch~8 & eight-stage $q=3329$ driver \\
    \bottomrule
  \end{tabular}
\end{table*}

The main contributions of this paper are as follows.
\begin{itemize}
  \item We construct an exact FP32 butterfly by splitting each twiddle into 8-bit and 4-bit parts, reducing both products separately, and combining only residues below $q$.
  \item We place two butterflies as four reduction wavefronts through row~46 of the 64-stage pipeline and fuse the fixed len-2 and len-4 stages through row~40.
  \item We limit ARM rearrangement to the fused output and use four CGLA scalar stores to form the next-stage 64-bit records across the remaining five stage transitions.
  \item All 105 FPGA runs match 817,152 coefficients.
  The batch-64 latencies are 27.5\,$\mu$s on FPGA and 6.74\,$\mu$s in the ASIC projection, and projected batch-8 system energy is 10.1\,$\mu$J per NTT.
\end{itemize}

The remainder of this paper is organized as follows.
Section~\ref{sec:related} reviews NTT accelerators and prior CGLA mappings.
Section~\ref{sec:mapping} presents the split-residue arithmetic and stage schedule, and Section~\ref{sec:experiments} reports correctness, timing, and energy results.
Section~\ref{sec:discussion} analyzes offload depth and the batch tradeoff before Section~\ref{sec:conclusion} concludes the paper.

\section{Related Work}
\label{sec:related}
\begin{figure}[t]
  \centering
  \includegraphics[width=\columnwidth]{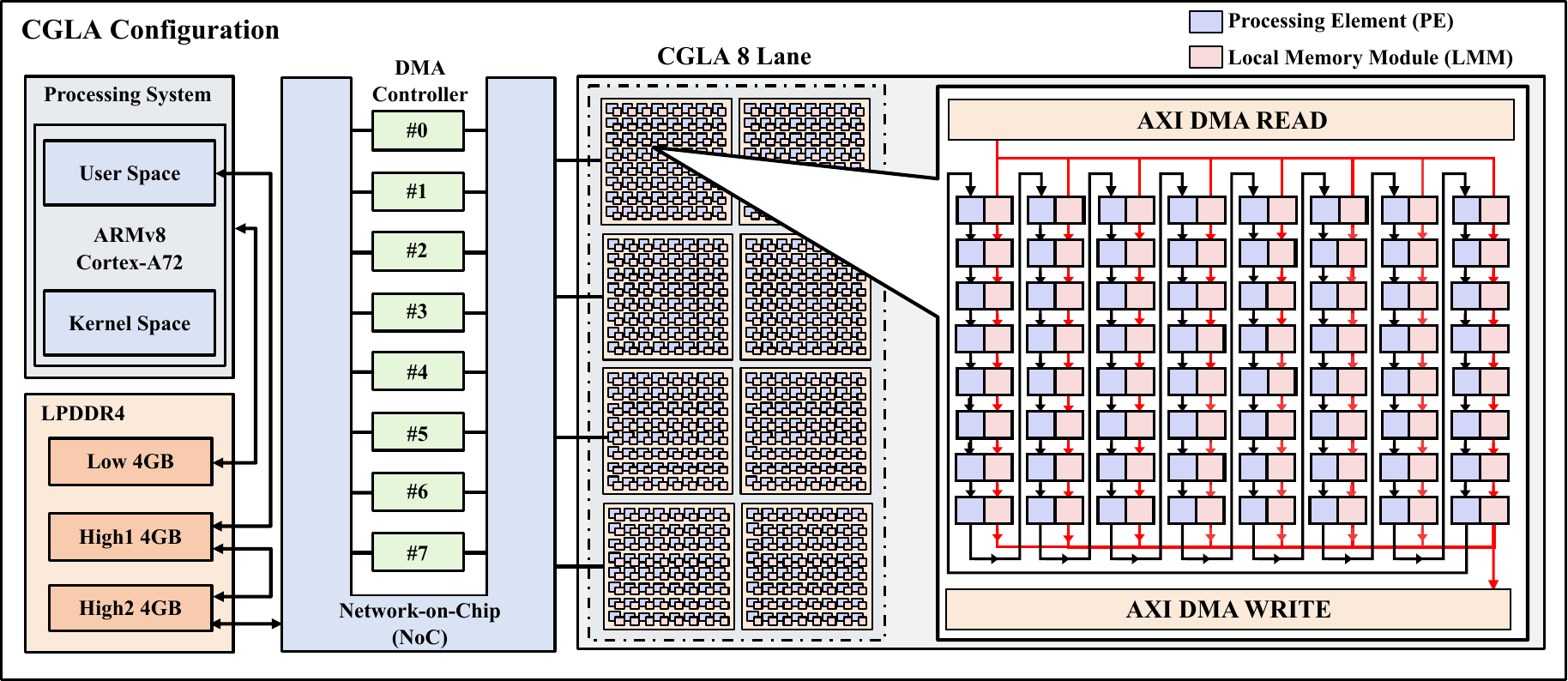}
  \caption{CGLA FPGA prototype used in this work.
  The PS contains the ARM host and DMA control, while the PL contains the CGLA array.}
  \label{fig:cgla-fpga-prototype}
\end{figure}
Existing NTT acceleration work can be separated by the execution substrate and the amount of arithmetic dedicated to the transform.
Transform-specific FPGA and ASIC accelerators specialize both arithmetic and memory systems for NTT execution~\cite{roy2023hardware}.
Saoudi et al.\ use a dedicated Kyber NTT datapath, while Nguyen et al.\ use a compact pipelined architecture~\cite{saoudi2024ntt,nguyen2024compact}.
Gao et al.\ emphasize throughput and BRAM efficiency with a streaming NTT/INTT datapath~\cite{gao2025bram}.
Sonbul et al.\ focus on pipelining and modular reduction, whereas Ni et al.\ remove BRAM from a lightweight NTT core~\cite{sonbul2025shiftadd,ni2023lightweight}.
Shrivastava et al.\ instead share hardware between FFT and NTT~\cite{shrivastava2025unified}.
Other designs target multi-scheme support, area--time product, throughput, or protocol integration~\cite{kundi2024mlkem,waris2025atp,bertels2025throughput,xing2021compact,nguyen2022areatime}.

In contrast, GPU and AI-accelerator implementations retain a programmable execution substrate but use platform-specific execution and memory interfaces.
Wan et al.\ and ConvKyber map Kyber operations to AI accelerators, while HI-Kyber uses GPU execution~\cite{wan2022kyberai,zhou2024convkyber,ji2023hikyber}.
Bao et al.\ place an NTT accelerator on the Versal AI Engine~\cite{bao2025versalntt}.
Zhu et al.\ report 10\,nJ for a 28\,nm computing-in-memory NTT accelerator with protocol-specific parallel circuits~\cite{zhu2026cim}.

Earlier CGLA studies map LLM inference~\cite{uetani2024cgla}, multi-dimensional U-Nets~\cite{duong2025unet}, and matrix-vector operations for approximate k-NN search~\cite{kim2025knn}.
Each study retains the same physical array and changes only the operation schedule and LMM dataflow.
However, these mappings are organized around matrix, convolution, or dot-product dataflow.
NTT instead requires an exact modular-reduction chain, stage-dependent coefficient pairing, and a packed-record layout updated at every stage boundary.
Cui et al.\ place a reconfigurable polynomial array inside a Kyber-specific FPGA controller~\cite{cui2025instruction}.
Our implementation uses an unchanged CGLA shared with non-cryptographic kernels and maps the exact butterfly and stage transitions through its existing operations and memory interface.

\section{Proposed Exact Full-Stage Mapping}
\label{sec:mapping}

\subsection{CGLA Architecture and Mapping Target}

CGLA is a programmable spatial architecture that arranges PEs and software-managed LMMs as an ordered linear pipeline~\cite{uetani2024cgla}.
The compiler targets an ordered sequence of logical rows instead of placing operations on a two-dimensional CGRA mesh.
It maps loop bodies sequentially to the logical rows without a per-kernel placement-and-routing search, shortening compilation~\cite{uetani2024cgla}.
Kernels change only the logical operation schedule, while the physical array remains fixed.
Compilation therefore resembles targeting a CPU instruction set rather than constructing a new FPGA datapath.
The 64 rows, supported operations, LMM capacity, and transfer bandwidth remain physical constraints.

Fig.~\ref{fig:cgla-fpga-prototype} shows the CGLA FPGA prototype used to evaluate the mapping.
The evaluated board combines a Processing System (PS), Programmable Logic (PL), Direct Memory Access (DMA), and external memory.
An ARM Cortex-A72 on the PS controls DMA, while the PL contains the 64-stage CGLA pipeline and its LMMs.
Each PE issues mapped \texttt{exe()} operations, and \texttt{mop(LDWR/STWR)} streams records through the LMM interface.
Here \texttt{LDWR} and \texttt{STWR} denote record load and store operations.
The arithmetic operations \texttt{OP\_FML} and \texttt{OP\_FAD} provide FP32 multiplication and addition.
The mapping uses \texttt{OP\_FMS}, a multiply-subtract operation, with a unit multiplier to form subtraction candidates.
\texttt{CMP\_GE} compares operands and \texttt{CMOV} conditionally selects a value.
The loaded configuration fixes the operation sequence across the logical rows, while loop iterations stream different record pairs through that sequence.
Row occupancy therefore represents the dependency depth of one mapped iteration rather than the number of butterflies in an NTT.
LMM supplies the coefficient and twiddle streams, and scalar stores return results to record addresses selected for the next stage.
Because end-to-end timing covers the PS--PL interface, the butterfly must fit the existing 64 rows and emit a layout that the following CGLA call can consume.

\begin{figure*}[t]
  \centering
  \includegraphics[width=0.80\textwidth]{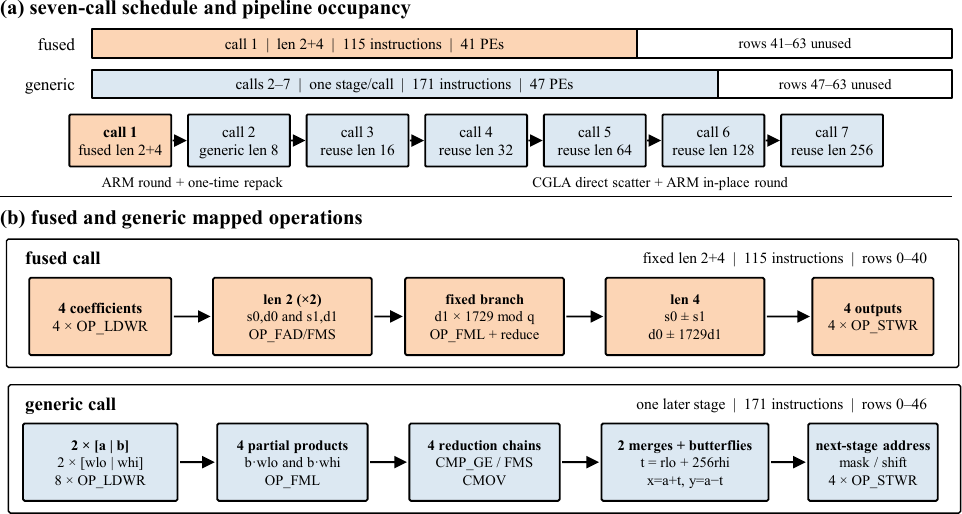}
  \caption{Exact seven-call CGLA mapping for the eight-stage driver.
  Panel (a) shows the fused call, six generic calls, and their 41/47-PE pipeline occupancy.
  Panel (b) expands the fused len-2/4 chain and the generic split-residue butterfly chain.}
  \label{fig:fullstage-mapping}
\end{figure*}

Fig.~\ref{fig:fullstage-mapping} summarizes the seven-call schedule and the fused and generic operation chains.
Panel~(a) separates the 41-PE fused len-2/4 call from the six 47-PE generic calls and marks configuration reuse for the five calls after len-8.
Panel~(b) expands the paths from record loads through multiplication, modular reduction, butterfly correction, and next-stage stores.

ML-KEM (FIPS~203) operates over $\mathbb{Z}_q[x]/(x^{256}+1)$ with $q=3329$ and uses a seven-layer incomplete negacyclic NTT~\cite{fips203}.
Montgomery-domain constants are an implementation representation choice, rather than a requirement of the transform specification.
We evaluate an eight-stage radix-2 $N=256$ driver over the same modulus and operand range to expose every stage transition of the mapped butterfly.
The forward driver applies a Cooley--Tukey butterfly over eight stages with a bit-reversal permutation.
At stage~$s$ ($s=0,\ldots,7$, $\textit{len}=2^{s+1}$), each butterfly computes
\begin{align}
  x &= a + w \cdot b \pmod{q}, \label{eq:bfx}\\
  y &= a - w \cdot b \pmod{q}, \label{eq:bfy}
\end{align}
where $w$ is a stage- and group-specific twiddle factor drawn from the power table of $\omega=17$, a primitive 256-th root of unity modulo $q$.
This full transform corresponds to cyclic convolution modulo $x^{256}-1$.
Adapting the implementation to FIPS~203 requires its twiddle order, seven layers terminating in coefficient pairs, and pairwise polynomial multiplication with the specified zeta factors.
The inverse transform also requires the inverse schedule and final scaling by $128^{-1}=3303\bmod q$.
Consequently, matching the modulus alone does not establish a standard-compliant ML-KEM NTT/INTT implementation.
All operands $a$, $b$, $w$ lie in $[0, q-1] = [0, 3328]$, requiring 12-bit representation.

\subsection{Mapping Requirements and Record Format}

The call interface packs two 32-bit FP32 bit patterns into each 64-bit coefficient record $[a_i\mid b_i]$.
A corresponding twiddle record contains $[w_{\mathrm{lo},i}\mid w_{\mathrm{hi},i}]$ for the same butterfly.
At a call boundary, each coefficient must represent an integer in $[0,q-1]$ so that the next multiplication and correction use the same bounds.
The arithmetic schedule must also end before row~64, while the output addresses must reflect the pair distance of the following stage.
For len-8 through len-256, stage-dependent twiddle records and offsets change without changing the generic arithmetic configuration.
The fused len-2/4 call uses a different output order and is the only boundary that requires host repacking.

\subsection{Exact Split-Residue Butterfly}

We split the twiddle into an 8-bit low part and a 4-bit high part and compute both partial products with FP32 \texttt{OP\_FML}.
\begin{equation}
  w = w_\mathrm{lo} + 256 w_\mathrm{hi},
  \quad w_\mathrm{lo}\in[0,255],\;w_\mathrm{hi}\in[0,13].
\end{equation}
CGLA then computes
\begin{equation}
  p_\mathrm{lo}=b w_\mathrm{lo}, \qquad
  p_\mathrm{hi}=b w_\mathrm{hi}.
\end{equation}
Product bounds are $p_\mathrm{lo}\leq848{,}640$ and $p_\mathrm{hi}\leq43{,}264$, both below $2^{24}$, so FP32 represents both integer products exactly.
On the FPGA prototype, \texttt{OP\_FAD} returned values one ULP below the exact result for some direct $p_\mathrm{lo}+256p_\mathrm{hi}$ additions despite the integer sum remaining below $2^{24}$.
ULP denotes one unit in the last place of the FP32 representation.
Correctly rounded IEEE FP32 arithmetic would represent this integer sum exactly, so the observation concerns the tested implementation rather than FP32 representability.
The responsible datapath behavior has not been isolated by an RTL-level analysis.
The mapping therefore computes
\begin{equation}
  r_\mathrm{lo}=p_\mathrm{lo}\bmod q,\quad
  r_\mathrm{hi}=256(p_\mathrm{hi}\bmod q)\bmod q,
\end{equation}
followed by $t=(r_\mathrm{lo}+r_\mathrm{hi})\bmod q$.
Because $bw=p_\mathrm{lo}+256p_\mathrm{hi}$, reducing the two terms separately preserves their sum modulo $q$.
Both residues are below $q$, so their integer sum is below $2q$ and is exactly representable in FP32.
This bound supports the arithmetic construction but does not prove that every prototype operation is correctly rounded.
Reduction uses descending powers of two times $q$ with \texttt{CMP\_GE}, \texttt{CMOV}, and \texttt{FMS}.
The low-product path uses eight steps from $2^7q$ to $q$, while the high-product path uses four steps before multiplication by 256 and the same eight-step reduction.
These step counts follow from $p_\mathrm{lo}<2^8q$, $p_\mathrm{hi}<2^4q$, and $256(p_\mathrm{hi}\bmod q)<2^8q$.
Each conditional subtraction computes the comparison and the \texttt{FMS} subtraction candidate in parallel, after which \texttt{CMOV} selects the source or candidate.
For positive FP32 values, the compared bit patterns preserve numeric order.
Given $a,t\in[0,q-1]$, $a+t$ is below $2q$ and $a-t$ lies between $-(q-1)$ and $q-1$.
One conditional subtraction for $x$ and one conditional addition for $y$ therefore return both outputs to $[0,q-1]$.
The corrected butterflies in Eqs.~\eqref{eq:bfx} and~\eqref{eq:bfy} use these bounds.
The operation sequence can be expressed with the following bounded reduction, where $\operatorname{select}$ maps to \texttt{CMOV}.
\begin{align*}
 R(v,h)&\colon\quad\text{for }j=h,h-1,\ldots,0,\\
 v&\leftarrow\operatorname{select}(v\geq 2^j q,\ v-2^j q,\ v),\\
 r_{\mathrm{lo}}&=R(bw_{\mathrm{lo}},7),\\
 r_{\mathrm{hi}}&=R(256R(bw_{\mathrm{hi}},3),7),\\
 t&=R(r_{\mathrm{lo}}+r_{\mathrm{hi}},0),\\
 x&=R(a+t,0),\\
 y&=\operatorname{select}(a\geq t,a-t,a-t+q).
\end{align*}
Two such butterflies are interleaved in the generic call before four scalar stores construct the next-stage records.
\subsection{Pipeline Schedule and Stage Fusion}

Direct-scatter addressing moves the two $a+q$ candidates ahead of the late reduction chain and removes redundant zero-add copies.
Compilation produces a 171-instruction generic pair-wavefront kernel through row~46 of the 64-stage pipeline.
Each pipeline iteration receives two coefficient records and two corresponding twiddle records in the format defined above.
The schedule preserves the dependencies of the two low- and two high-residue chains and issues their \texttt{CMP\_GE}, \texttt{FMS}, and \texttt{CMOV} operations as wavefronts.
Here a wavefront is a set of independent operations progressing through successive mapped rows while preserving each chain's dependencies.
Interleaving the four independent residue chains fills dependency gaps without changing their arithmetic order.
After both residues form $t_0$ and $t_1$, the two $x/y$ correction pairs remain independent.
Each iteration processes two butterflies while successive record pairs stream through the fixed row mapping.
Mask and shift operations derive the next-stage record and lane offsets from the current byte offset and stage half-length before the arithmetic chain completes.
Four scalar stores then write the corrected $x_0$, $y_0$, $x_1$, and $y_1$ values directly into that layout.
Fixed twiddles $\{1,1,1,1729\}$ allow the first two stages to execute len-2 and len-4 in one 115-instruction fused call through row~40.
Its two len-2 butterflies form $s_0,d_0,s_1,d_1$.
Only the $d_1$ branch requires multiplication by 1729 before the len-4 combination and four output stores.
This fusion removes the call boundary and intermediate record transfer between len-2 and len-4.
Stage-ordered twiddle records and stage-dependent offsets allow one 171-instruction generic configuration to process each stage from len-8 through len-256.
One fused call is followed by six generic calls, of which the five after len-8 reuse the configuration loaded by the len-8 call.

\subsection{Stage Chain and Host Boundary}
CGLA outputs at each stage boundary are integer-near FP32 values.
Feeding their raw bit patterns to the next stage caused errors, so the host applies \texttt{vrndn\_f32} to both lanes of each record.
The host normalization converts these values to integer-valued FP32 records but does not repeat the modular multiplication or reduction.
ARM first rounds and repacks the fused output once into the len-8 $[a\mid b]$ layout.
At the five subsequent generic transitions, CGLA stores each output directly in the next stage's layout.
The host therefore only rounds records in place without rearranging them or materializing an integer polynomial.
Direct stage chaining thus removes generic-stage record rearrangement while retaining host normalization and a host-visible call boundary at every stage.
Twiddle records are stored in stage order, and a 10-bit byte-offset mask lets the generic kernel reuse one 128-record template across all batch items.

\begin{figure*}[t]
  \centering
  \includegraphics[width=\textwidth]{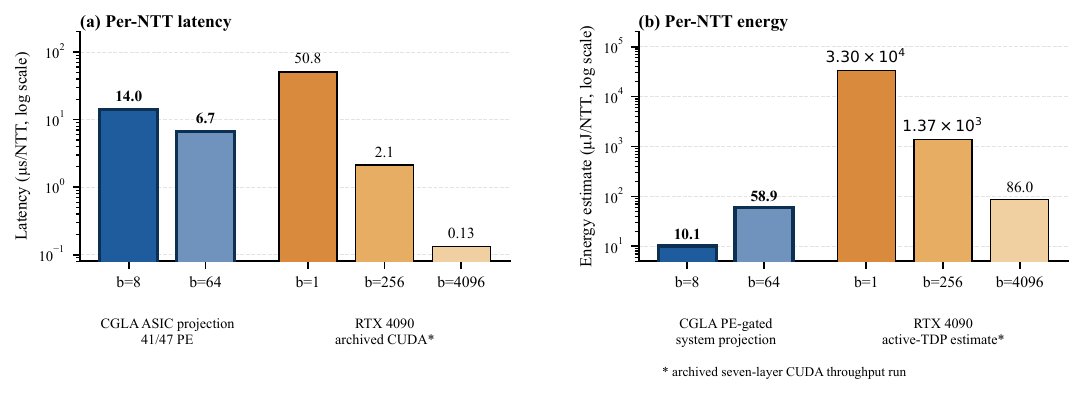}
  \caption{Contextual CGLA projections and archived RTX~4090 throughput results for different transforms.
  Time per NTT is batch-normalized and does not represent single-request response latency.
  CGLA uses the eight-stage driver, while RTX uses seven-layer CUDA timing and active-TDP energy estimates.}
  \label{fig:platform-context}
\end{figure*}

\begin{figure*}[t]
  \centering
  \includegraphics[width=\textwidth]{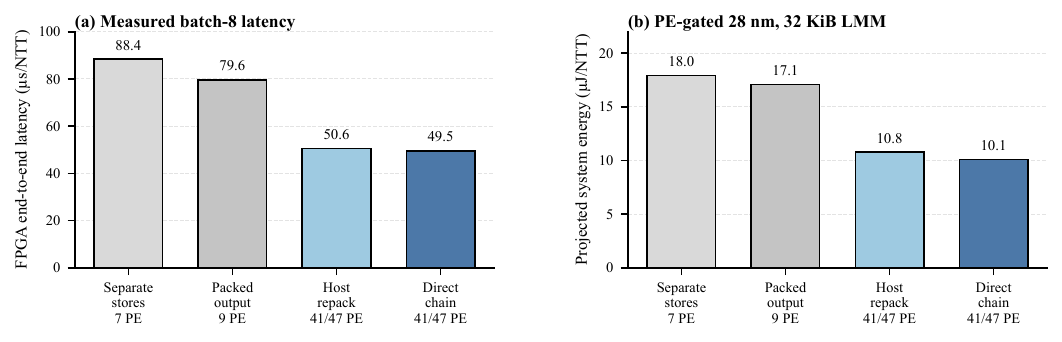}
  \caption{Exact batch-8 offload-depth ablation over 21 runs per implementation.
  Panel (a) reports measured FPGA end-to-end latency, and panel (b) reports PE-gated 28-nm system-energy projections with a fixed 32-KB LMM configuration.
  The 7/9-PE implementations and the host-repack/direct-chain variants form two same-binary control pairs.}
  \label{fig:offload_ablation}
\end{figure*}
\section{Experiments and Results}
\label{sec:experiments}
The experiments examine correctness, end-to-end timing, projected energy, and the effect of moving additional stage work from the host to CGLA.
For the offload-depth comparison, the FPGA clock, batch, inputs, and end-to-end timing boundary remain fixed.
Every run is checked coefficient by coefficient against the CPU reference.
Fig.~\ref{fig:platform-context} places the measured and projected results in their platform context, while Fig.~\ref{fig:offload_ablation} isolates offload depth at batch~8.

\subsection{Experimental Setup and Correctness}

We implement the full-stage driver in C on the 145-MHz, 64-PE CGLA FPGA prototype on an AMD Versal VPK180 with up to 512~KB LMM.
An ARM Cortex-A72 at up to 1.4\,GHz controls the PS side.
FPGA end-to-end timing includes ARM preparation, CGLA transfer and execution, and output handling.
Each result uses 21 runs, and the batch-8 and batch-64 energy cohorts record call phases.
The batch-64 latency cohort is unprofiled, while the ablation uses 21 batch-8 runs per implementation.
Run-level medians include every retained trial, including slow trials.

Following prior CGLA evaluations, the phase-based model uses CGLA power estimates from synthesis with Synopsys Design Compiler and a TSMC 28-nm standard-cell library.
For batch $B$, the fused call uses $2B$~KB, while a generic call uses $2B+1$~KB including the shared twiddle template.
Batch~8 uses a 17-KB live set in 32~KB LMM, with fused and generic synthesis estimates of 1.18 and 1.35\,W.
Batch~64 uses a 129-KB live set in 256~KB LMM, with corresponding estimates of 20.5 and 23.4\,W.
These estimates come from different LMM-capacity synthesis points, with unchanged mapped PE counts.
A component-level memory-power and switching-activity breakdown is unavailable, so the large power increase cannot be attributed to a verified circuit mechanism.
The resulting batch-energy ordering is conditional on these synthesis-table values.
For the Cortex-A72, the model uses 0.649\,W during host processing and 0.249\,W idle power while CGLA is active~\cite{Humrick_CortexA72_2016}.

Correctness is checked against a pure-C Cooley--Tukey driver with Barrett reduction (\texttt{ntt\_mlkem.c}).
We evaluate the final 41/47-PE implementation in 105 runs comprising 21 batch-64 latency runs, 21 batch-64 profile runs, 21 batch-8 direct-chain latency runs, and 42 alternating direct-chain/host-repack profile runs.
The four cohorts match all 817,152 coefficients with the CPU reference.
A host oracle also checks all 11,082,241 split-product cases and all 11,082,241 $(a,t)$ correction pairs.
These exhaustive host arithmetic checks validate the bounded residue construction, while the FPGA runs test the mapped execution on the retained input cohorts.
They do not constitute an exhaustive hardware proof for every transform input or for the prototype rounding behavior.

\subsection{Measured Timing and ASIC Projection}

Static timing analysis under the same 28-nm synthesis flow reports an 840-MHz CGLA clock.
Following the nominal convention used in prior CGLA evaluations, the projection is $T_\mathrm{ARM}+T_\mathrm{CGLA}/6$.
Here $T_\mathrm{CGLA}$ is the measured non-ARM interval, including configuration, transfer, execution, and drain, whereas ARM preparation and normalization remain unscaled.
This is an optimistic all-phase scaling scenario because DMA and configuration bandwidth need not increase with the PE clock.
A phase-specific alternative retains those intervals and scales only execution, giving $T_\mathrm{ARM}+T_\mathrm{CONF}+T_\mathrm{DMA}+T_\mathrm{DRAIN}+T_\mathrm{OTHER}+T_\mathrm{EXEC}/6$.
The reported 6.74\,$\mu$s value uses the original all-phase scenario, not this execution-only alternative.
Table~\ref{tab:phases} separates the archived batch-64 timing counters.
Applying execution-only scaling to each retained run gives a median of 15.87\,$\mu$s per NTT, illustrating the dependence on interface scaling assumptions.
\begin{table}[t]
\centering
\caption{Batch-64 FPGA phase medians in $\mu$s per NTT over 21 runs. Phase medians need not sum to the median total.}
\label{tab:phases}
\small
\begin{tabular}{lr}
\toprule
Measured interval & Time \\
\midrule
ARM preparation and normalization & 2.594 \\
Configuration & 1.141 \\
Input loading & 4.719 \\
Array execution & 13.906 \\
Output drain & 4.797 \\
Other setup and unassigned time & 0.328 \\
End-to-end total & 27.469 \\
\bottomrule
\end{tabular}
\end{table}
For the retained batch-64 runs, median FPGA end-to-end latency is 1.76\,ms per group, or 27.5\,$\mu$s per NTT, with an interquartile range of 27.4--27.5\,$\mu$s.
Median CGLA-side time is 1.59\,ms per group, or 24.9\,$\mu$s per NTT, and the ASIC projection is 6.74\,$\mu$s per NTT with in-place NEON normalization retained in $T_\mathrm{ARM}$.

\subsection{Energy Results and Platform Context}

Both batch configurations use the same ASIC projection.

At batch~8, the median \texttt{EXEC} and remaining CGLA-side intervals are 129 and 212\,$\mu$s per group.
After the same timing projection, using 1.35\,W during \texttt{EXEC} and 0.0606\,W outside \texttt{EXEC} gives a PE-gated CGLA-only median of 3.90\,$\mu$J per NTT.
Adding the phase-accounted Cortex-A72 energy gives a system-energy median of 10.1\,$\mu$J per NTT.
For batch~64, the corresponding \texttt{EXEC} and non-\texttt{EXEC} medians are 887 and 701\,$\mu$s per group.
With 256~KB LMM, PE-gated CGLA-only and system medians are 55.8 and 58.9\,$\mu$J per NTT.
Among the tested batch sizes, batch~8 minimizes energy and batch~64 minimizes measured latency for the 41/47-PE implementation.

Fig.~\ref{fig:platform-context} compares CGLA at batches 8 and 64 with RTX~4090 results at batches 1, 256, and 4096.
The RTX energy estimates use nominal CPU and GPU TDP values~\cite{nvidia_ada,Intel_Xeon_w5-2455X_Specs}.
Under the stated accounting, the projected CGLA batch-8 system energy is 10.1\,$\mu$J per NTT, compared with 86.0\,$\mu$J for the RTX batch-4096 active-TDP estimate.
RTX batch~4096 reaches 0.13\,$\mu$s per NTT, compared with a projected 6.7\,$\mu$s for CGLA batch~64.
CGLA batch~8 minimizes energy within the tested CGLA projections, while the archived RTX batch~4096 point has the smallest amortized time.
The GPU values originate from the authors' archived seven-layer CUDA benchmark, whose timing records lack a same-run device identity and build hash.
That program skips correctness checking, so the GPU rows provide historical throughput context rather than a matched-transform performance or energy comparison.
Dedicated-hardware values in Table~\ref{tab:design_space} likewise use their original transform, batching, and energy boundaries.
Standardized seven-layer NTT/INTT execution on CGLA remains outside the measured scope.

\section{Discussion}
\label{sec:discussion}

To examine how output packing and stage-to-stage data placement affect latency and energy in the CGLA implementation, we compare four exact batch-8 implementations.
Fig.~\ref{fig:offload_ablation} applies the same FPGA end-to-end protocol and toolchain to all four implementations.
A 7-PE kernel receives a host-computed $t=bw\bmod q$ and stores $x$ and $y$ separately, while the 9-PE kernel packs both outputs into one store.
Both 41/47-PE variants use identical fused and generic kernels and differ only in ARM rearrangement of the five generic-stage outputs.
Each implementation uses eight deterministic inputs per run, and all 84 runs match 2,048 output coefficients with the CPU reference.
Median FPGA latencies are 88.4, 79.6, 50.6, and 49.5\,$\mu$s per NTT for 7 PE, 9 PE, host repacking, and direct stage chaining, respectively.

The 7- and 9-PE pair isolates output packing while retaining the host twiddle product.
The two 41/47-PE variants isolate direct scatter from host repacking.
Compared with host repacking in Fig.~\ref{fig:offload_ablation}, direct scatter reduces median latency by 1.0\,$\mu$s and median energy by 0.66\,$\mu$J per NTT, with lower latency in 17 of 21 paired runs.
The 37.8\% gap also includes CGLA-side multiplication, reduction, first-two-stage fusion, and stage chaining, so it reflects aggregate offload depth rather than PE count alone.

The batch sweep exposes a second tradeoff because latency and energy respond differently to LMM capacity, PE count, and the amortization of configuration and transfer.
Batch~8 uses 32~KB LMM and gives the lowest tested system-energy projection of 10.1\,$\mu$J per NTT.
Batch~64 amortizes configuration and transfer to 6.74\,$\mu$s per NTT, but its 256-KB LMM raises system energy to 58.9\,$\mu$J per NTT.
Direct scatter removes record rearrangement between generic calls.
A fully CGLA-resident chain still requires exact normalization because the tested CGLA rounding instructions did not match the full-stage oracle.
The energy comparison is limited to the CGLA ASIC projection and nominal CPU/GPU TDP values because direct power was not measured.
The mapping occupies rows 0--40 for the fused call and 0--46 for the generic call within the 64-row array.
These row spans describe scheduling resources, rather than a measured area or power saving from avoiding a dedicated NTT unit.
Matched CGLA implementations of Barrett and Montgomery reduction have not been evaluated under this final full-stage protocol.
Accordingly, the conditional-subtraction mapping demonstrates feasibility without establishing optimality against those alternatives or an integer-multiplier target.
The fixed mapped reduction steps do not require a data-dependent iteration count, but host control, rounding, and memory behavior have not undergone a constant-time audit.
No side-channel resistance or complete ML-KEM timing guarantee follows from these measurements.
Other moduli require new operand bounds, reduction schedules, and hardware checks before the same construction can be applied.

\section{Conclusion}
\label{sec:conclusion}

This work mapped an exact eight-stage $N=256$ NTT driver over the ML-KEM modulus to a programmable CGLA through split-residue FP32 arithmetic and a seven-call 41/47-PE stage chain.
The evaluated transform is cyclic and retains host normalization at stage boundaries.
The standardized incomplete negacyclic NTT/INTT and complete ML-KEM execution remain unimplemented on the evaluated CGLA path.
Across four cohorts, all 105 FPGA hardware runs matched 817,152 output coefficients.
At batch~64, measured FPGA latency is 27.5\,$\mu$s per NTT and projected ASIC latency is 6.74\,$\mu$s per NTT.
The PE-gated system-energy projection is 10.1\,$\mu$J at batch~8 and 58.9\,$\mu$J at batch~64.
These values reflect the latency--energy tradeoff introduced by the larger LMM configuration.
The next step is to apply the measured schedule to the seven-layer FIPS~203 NTT/INTT pipeline and evaluate fixed-LMM capacity under the high-throughput configuration.

{\fontsize{7.5}{8.1}\selectfont
\interlinepenalty=10000
\bibliographystyle{IEEEtran}
\bibliography{IEEEabrv,references}
}

\end{document}